\newcommand{\D}{\Delta}

\newcommand{\w}{\omega}
\newcommand{\nn}{\nonumber}

\newcommand{\eps}{\varepsilon}
\newcommand{\al}{\alpha}
\newcommand{\jd}{\gamma_d}
\newcommand{\js}{\gamma_s}
\newcommand{\jn}{\gamma_n}

\documentclass[a4paper,twocolumn,pra,showpacs]{revtex4}

\usepackage{amsmath}
\usepackage{graphicx}
\usepackage{subfigure}
\newcommand{\curdir}{.}
\newcommand{\pics}{\curdir/pics}

\begin{document}

\title{A route to robust double pulse excitability in optically injected semiconductor lasers}
\author{Sergey Melnik}
\affiliation{Tyndall National Institute, Lee Maltings, Cork, Ireland}
\affiliation{Department of Applied Physics and Instrumentation, Cork Institute of Technology, Cork, Ireland}
\author{Oleg Rasskazov}
\affiliation{Tyndall National Institute, Lee Maltings, Cork, Ireland}
\affiliation{Department of Applied Physics and Instrumentation, Cork Institute of Technology, Cork, Ireland}
\author{Guillaume Huyet}
\affiliation{Tyndall National Institute, Lee Maltings, Cork, Ireland}
\affiliation{Department of Applied Physics and Instrumentation, Cork Institute of Technology, Cork, Ireland}


\begin{abstract}
We present and analyse a three-dimensional model for a quantum dot semiconductor laser with optical injection. This model describes recent experimental single and double excitable intensity pulses, which are related to a central saddle-node homoclinic bifurcation as in the Adler equation. Double pulses are related to a period doubling bifurcation and occur on the same homoclinic curve as single pulses. The bifurcation scenario consolidating single and double excitable pulses is described in detail.
\end{abstract}


\maketitle

\section{Introduction}
Semiconductor lasers with optical injection have been the subject of numerous studies to generate optical bistability~\cite{lugiato84}, increase the modulation bandwidth of directly modulated lasers~\cite{chrostowski06}, control the transverse mode structure~\cite{Li94} or analyse the appearance of spatial solitons~\cite{barland02}. These experiments have also revealed a rich and complicated dynamical behaviour that has motivated many studies~\cite{Henry85,Solari94,Erneux96,Alsing96, Wieczorek02,Krauskopf03,Wieczorek05}.

In its simplest form a semiconductor laser with optical injection constitutes an example of a forced non-linear oscillator which in addition to synchronisation can also display multistability, exhibit instabilities, cascades of bifurcations~\cite{Yeung98,Hohl99} and sudden chaotic transitions~\cite{wieczorek01}. It was also considered as a prototype for excitability since the phase dynamics can be described by the Adler equation~\cite{adler73} which was initially introduced to model synchronisation in electrical oscillators. In this equation the locking/unlocking transition is associated with a saddle-node bifurcation on a limit cycle, and as a result two different types of transient can be observed depending on the initial condition, which in the presence of noise gives rise to excitable pulses. Further theoretical studies of semiconductor lasers with optical injection have also predicted the appearance of multipulse excitability associated with n-homoclinic bifurcations~\cite{Wieczorek05,Wieczorek02,Krauskopf03}.

Although excitability was experimentally observed in several laser systems~\cite{plaza97,giudici97,wunsche02}, it was found only recently in semiconductor lasers with optical injection~\cite{Goulding07}. The experiment was carried out using a quantum dot laser (QDL), which displayed strongly damped relaxation oscillations~\cite{obrien04} and relatively low $\al$-factor~\cite{muszalski04}. Single and double excitable pulses were observed for a wide range of parameters. 

\begin{figure}[t!]
        \includegraphics[width=0.95\columnwidth]{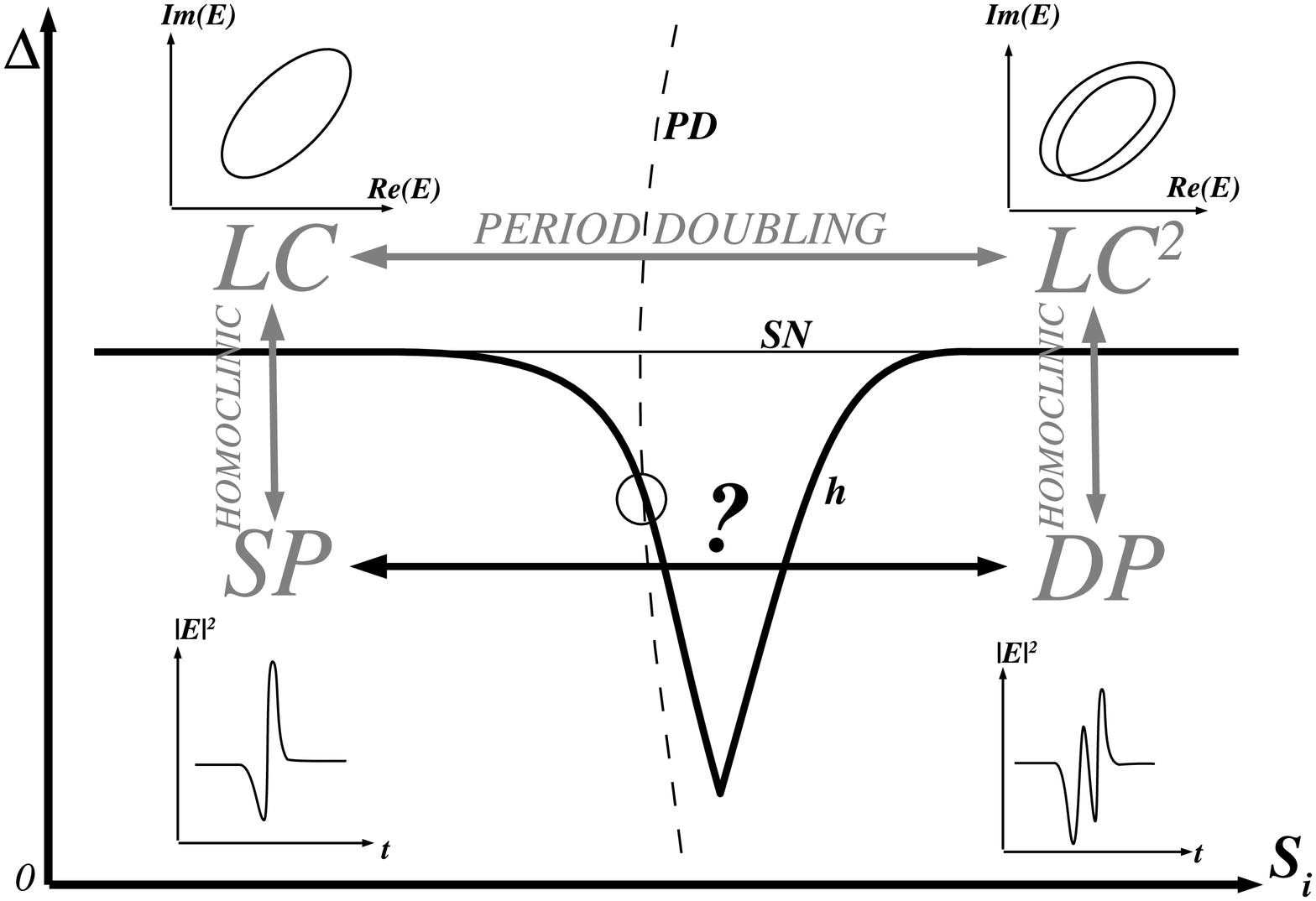}
\caption{Mechanism for single and double pulse excitability. $SN$ is the saddle-node line, $PD$ is the period doubling curve and $h$ is the homoclinic curve. The region $LC$ contains stable limit cycle and the region $LC^2$ contains stable period-2 limit cycle. Single and double pulses are found in the regions $SP$ and $DP$ correspondingly, the transition between these regions is the main point of this paper.
}
\label{PD_HOM}	
\end{figure}

Similar behaviour was also found theoretically in a four-dimensional dynamical system~\cite{Goulding07}, where the presence of double pulses was related to a period doubling bifurcation `crossing'~\footnote{Period doubling cannot happen simultaneously with a homoclinic bifurcation. This means that when period doubling and homoclinic bifurcation curves `cross' each other in the parameter space, these bifurcations happen in fact with different limit cycles in the phase space.} a homoclinic curve as shown in Fig.~\ref{PD_HOM}. This figure is a bifurcation diagram sketch which contains a saddle-node line $SN$, a period doubling curve $PD$ and a homoclinic curve $h$. The 'homoclinic tooth', defined as the part of $h$ which does not coincide with $SN$, plays a major role in the transition between single and double pulses. On the left of the tooth, the limit cycle $LC$ undergoes a homoclinic bifurcation on $SN$, leading to a formation of a stable and an unstable equilibria connected by long and short heteroclines. Here, noise can induce the system to jump from the stable equilibrium to the vicinity of the unstable one, which results in a phase space excursion along the long heteroclinic trajectory that is shaped like a `just broken' limit cycle. On the right of the tooth, the period-2 limit cycle $LC^2$ undergoes a homoclinic bifurcation on the same line $SN$, leading to the appearance of double excitable pulses $DP$ in the presence of noise.

This explanation differs from the one predicted earlier using conventional laser rate equations with optical injection ~\cite{Wieczorek05,Wieczorek02,Krauskopf03}, where the multipulse excitability was predicted inside the main homoclinic tooth. The distinguishing feature of our `period doubling' excitability mechanism in comparison with the n-homoclinic scenario is that the regions in $(S_i,\D)$ parameter space where single and double pulses appear are separated by the tooth and are sufficiently large to make experimental observation possible.

Above the $SN$ line the transition between the limit cycle $LC$ and the period-2 limit cycle $LC^2$ is clearly defined by the period doubling bifurcation curve $PD$. There is, however, a continuous transformation between $LC$ and $LC^2$ along the homoclinic tooth and the period doubling curve $PD$ does not play a role in this transformation. Since a pulse resembles the shape of a `just broken' limit cycle, there is no definite boundary between single and double pulses.

This paper aims to describe the intricate bifurcation picture near the homoclinic tooth in order to understand the transition between single and double pulses. In the next section we introduce a simplified three-dimensional model which we use for the analysis. The third section describes the bifurcations of equilibria and limit cycles of the model as well as mentions the types of bistability, which were observed experimentally.  Finally, section four focuses on the bifurcations near the homoclinic tooth, which are important to understand the transition from single to double excitable pulses.

\section{Model}
In order to describe the dynamical behaviour of a QDL with optical injection we use the following three-dimensional model for the dimensionless complex electric field $E$ and the carrier population $N$ (the formulation as well as the values of parameters are given in the Appendix):
\begin{eqnarray}
\nn \dot E &=& \frac12 \left(g\frac{N -1}{N + |E|^2 +1} - 1 \right) (1+\imath \al )E+\sqrt{S_{i}}+\imath\D E,\\
\dot N &=& \eta \left(- N + P - B N\frac{|E|^2 +2}{N + |E|^2 +1}\right). 
\label{SMrev}
\end{eqnarray}
The two main parameters are the injection strength $S_i$ and the frequency detuning $\D$. We define detuning $\D$ as the difference between the frequencies of slave and master lasers. Positive detuning means that the wavelength of the master laser is larger than that of the free running slave laser. Since $S_i$ and $\D$ can be easily changed in an experiment, these parameters constitute our bifurcation parameter space. The parameters which describe the material properties are the differential gain factor $g$, the linewidth enhancement factor $\al$, carrier to photon decay rate ratio $\eta$, capture coefficient $B$, and the laser pump parameter is $P$.

When control parameters $S_i$ and $\D$ are varied, the number and stability of the solutions of the dynamical system (\ref{SMrev}) generally change as well. The next section will discuss bifurcations of equilibria and limit cycles. The bifurcations presented in this paper were obtained with AUTO2000 \cite{AUTO2000}, a software package for continuation and bifurcation problems in ordinary differential equations.

\section{Equilibria and limit cycles}
\begin{figure}[t!]
        \includegraphics[width=0.95\columnwidth]{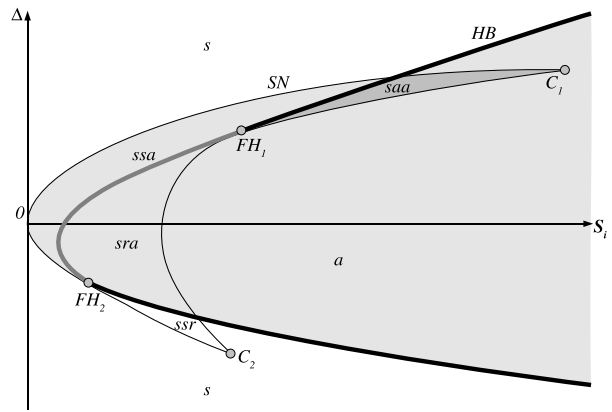}
\caption{Qualitative sketch of $(S_i,\D)$ parameter space with symbols indicating the stability of equilibria, where $a$ stands for a stable equilibrium (attractor), $r$ for a totally unstable equilibrium (repeller), and $s$ for a saddle equilibrium. The shaded region is the locking region and the dark shaded region is the region of bistability between two locked states.}
\label{SN_HB}
\end{figure}
The dynamical system (\ref{SMrev}) can have from 1 to 3 equilibria depending on the region in the $(S_i,\D)$ parameter space. Fig.~\ref{SN_HB} schematically represents equilibria bifurcation curves in the $(S_i,\D)$ parameter space. The thin curve $SN$ is a saddle-node (also known as fold or limit point) bifurcation curve along which two new equilibria appear. The bold line $HB$ is an (Andronov-)Hopf bifurcation curve along which limit cycles are created.

Since the system~(\ref{SMrev}) is in the frequency frame of the master laser, a stable equilibrium corresponds to a locked operation of the laser. In the locked regime the laser operates at constant power and at the injected master light frequency. Locking can occur in regions with at least one stable equilibrium, i.e. in the regions $ssa,sra,saa$, and $a$.

As in the experiment~\cite{Goulding07}, two types of bistability can occur in the dynamical system (\ref{SMrev}). Bistability between two stable equilibria is found in the dark shaded region $saa$ in Fig.~\ref{SN_HB}. As we leave the region $saa$ via the $HB$ curve, one of the stable equilibria bifurcates into a stable limit cycle. Thus, the second type of bistability between a limit cycle and an equilibrium can be found in the region $ssa$ above the black part of the $HB$ curve.

There are also four codimention-2 bifurcation points. Points $C_1$ and $C_2$ are cusp bifurcations, where three equilibria bifurcate, and imply the presence of hysteresis (see section 8.2.2 in~\cite{Kuznetsov98}). Points $FH_1$ and $FH_2$ are the codimension-2 fold-Hopf bifurcations, where the saddle-node and the Hopf curves are tangent (details on these and other bifurcations can be found in~\cite{Kuznetsov98}). Using the classifications from~\cite{Kuznetsov98}, $FH_1$ is a type 3 fold-Hopf bifurcation and $FH_2$ is a type 4 fold-Hopf bifurcation with reversed time.

\begin{figure}[t!]
\centering
        \includegraphics[width=0.95\columnwidth]{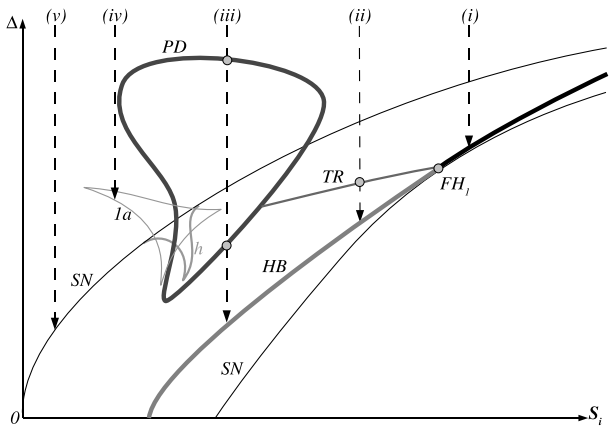}
\caption{The five different scenarios of disappearance of limit cycle $LC$. The torus bifurcation line $TR$ starts at the fold-Hopf point $FH_1$ and connects to the period doubling curve $PD$.}
\label{5WaysLC}
\end{figure}

The $HB$ curve can be divided into sub- and supercritical parts. Above $FH_1$ and below $FH_2$ the Hopf bifurcation curve is supercritical (shown in black in Fig.~\ref{SN_HB}) leading to formation of a stable limit cycle. Between $FH_1$ and $FH_2$ the Hopf bifurcation is subcritical (shown in gray in Fig.~\ref{SN_HB}) and therefore the created limit cycle is unstable. The stable limit cycle $LC$ which exists for large values of detuning $\D$ can disappear in five different ways depending on the value of $S_i$ as shown in Fig.~\ref{5WaysLC}:

(i) For $S_i$ above point $FH_1$, $LC$ disappears via a supercritical Hopf bifurcation on $HB$.

(ii) For $S_i$ just below point $FH_1$, $LC$ first loses its stability on a torus bifurcation curve $TR$ (which emanates from point $FH_1$) and subsequently disappears via a subcritical Hopf bifurcation on $HB$.

(iii) For even smaller values of $S_i$, $LC$ undergoes a period doubling bifurcation on the $PD$ curve before disappearing via a subcritical Hopf bifurcation on $HB$. 

(iv) The limit cycle $LC$ can also disappear via a limit cycle fold bifurcation.

(v) For very small values of $S_i$ the $LC$ undergoes a central saddle-node homoclinic bifurcation on the $SN$ curve, thus giving rise to single excitable pulses.

Most interestingly, for the case where $LC$ undergoes a period doubling bifurcation on the $PD$ curve, the stable newly formed period-2 limit cycle $LC^2$ can undergo a homoclinic bifurcation giving rise to double excitable pulses.

\begin{figure}[t]
\centering
        \includegraphics[angle=-90,width=0.95\columnwidth]{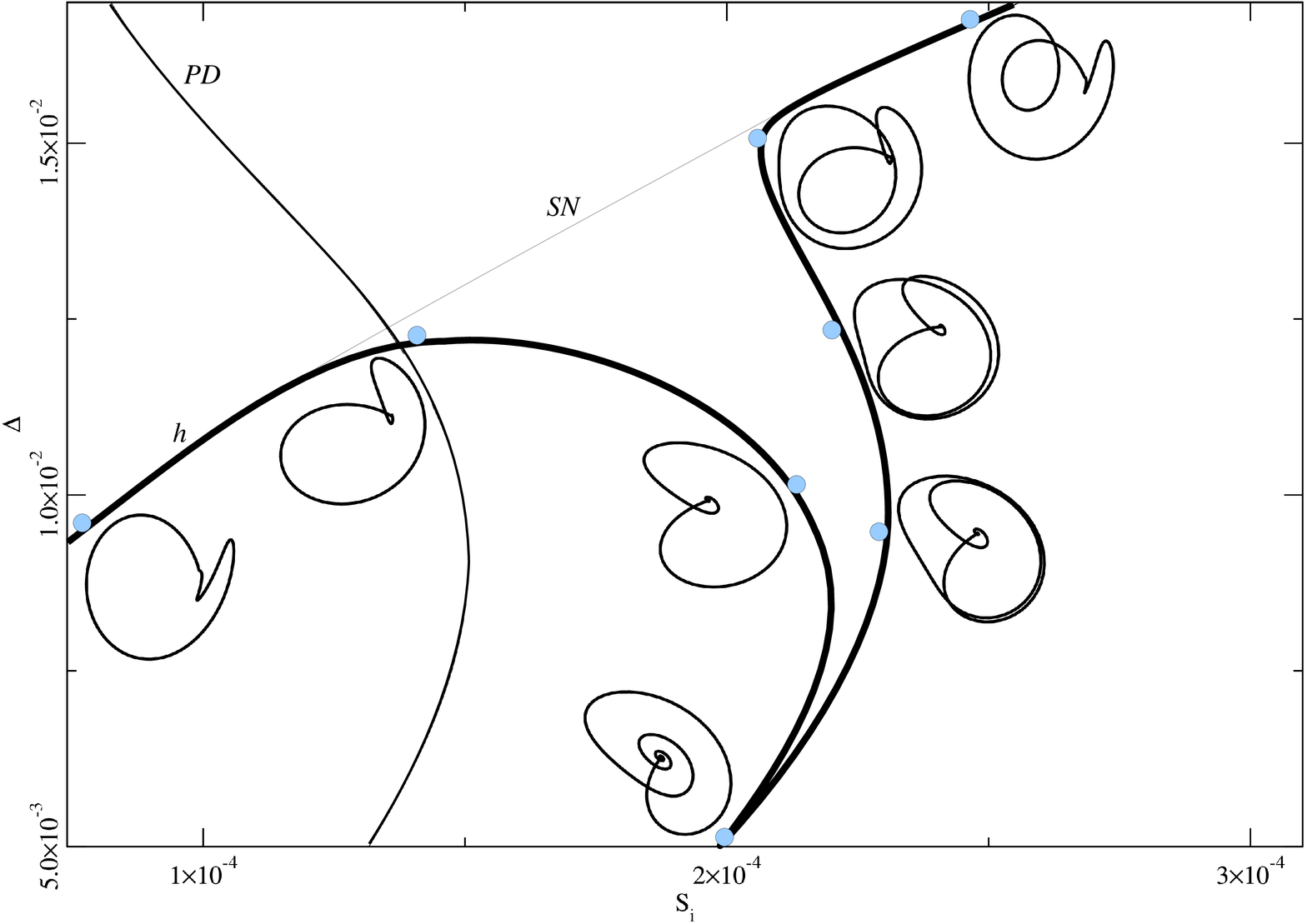}
\caption{Continuous transformation between the limit cycles $LC$ and $LC^2$ along the homoclinic tooth. $LC$ gradually winds up and becomes $LC^2$. Limit cycles are projected onto the plane $N=0$.}
\label{CAT}
\end{figure}

Let us first notice that on the left of the tooth the original $LC$ becomes homoclinic on the $SN$ curve. However on the right of the tooth (where the original $LC$ suddenly creates a period-2 cycle $LC^2$ while passing through $PDa$ curve) the $LC^2$ becomes homoclinic on the same curve $SN$ while the original $LC$ disappears via a subcritical Hopf bifurcation on $HB$. On the other hand, if the original $LC$ is followed in ($S_i,\D$) space very close along the homoclinic curve from the left side of the tooth to the right side, it will gradually wind up and become the period-2 cycle $LC^2$ as shown in Fig.~\ref{CAT}. Thus, the original $LC$ and the period doubled $LC^2$ are in fact the same object.

\section{Transition from single to double pulses}
\begin{figure}[b!]
\centering
        \includegraphics[width=0.95\columnwidth]{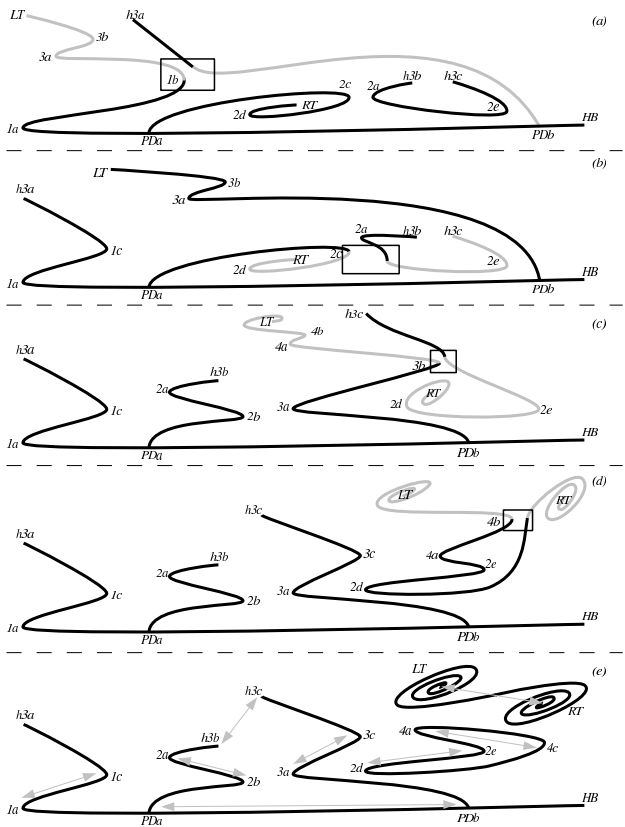}
\caption{Panels showing topological 1-parameter bifurcation diagrams of limit cycles around the tooth as $\D$ decreases. See Fig.~\ref{th} for the values of $\D$ corresponding to each panel as well as for notation. The curves represent maxima of limit cycles versus $S_i$.
}
\label{steps}
\end{figure}

\begin{figure*}[t!]
\centering
        \includegraphics[width=2\columnwidth]{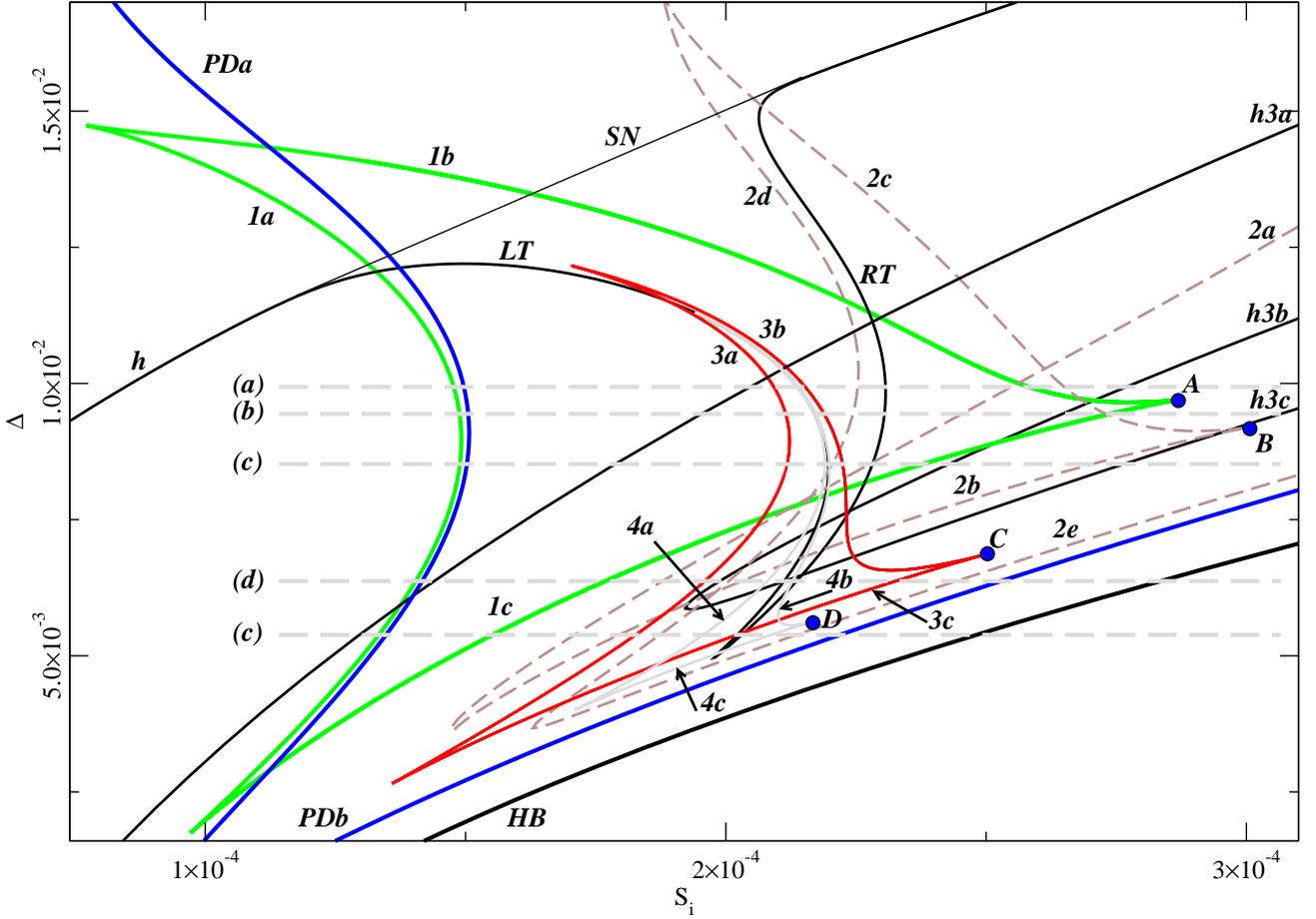}
\caption{Main bifurcation curves in ($S_i$,$\D$) parameter space near the homoclinic tooth. Dashed horizontal lines indicate the values of $\D$ for which 1-parameter bifurcation sketches were made in Fig.~\ref{steps}. $LT$ and $RT$ are the left and the right sides of the homoclinic tooth. $h3a$, $h3b$, $h3c$ are the different parts of another homoclinic curve. $HB$ is the Hopf bifurcation curve. $PDa$, $PDb$ are the two sides of the period doubling curve $PD$. Number labels mark the limit cycle fold bifurcation curves and the letters refer to the sides of that particular fold curve.
}
\label{th}
\end{figure*}

This section describes the sequence of bifurcations which enables transition from single to double pulses. The main steps as we move down to the tip of the homoclinic tooth are presented in Fig.~\ref{steps}. The panels show qualitatively different 1-parameter bifurcation sketches versus $S_i$, for the fixed values of $\D$ indicated in Fig.~\ref{th}.
On panel (a) the limit cycle is created from a homoclinic bifurcation at the left side of the tooth $LT$. Several fold bifurcations occur in 3b, 3a, 1b, 1a leading to the creation and destruction of several limit cycles. In $PDa$ period doubling bifurcation occurs. The period-2 limit cycle disappears in 2c via a fold bifurcation with one of the limit cycles created in the fold bifurcation 2d. The second limit cycle created in the fold bifurcation 2d undergoes homoclinic bifurcation at the right side of the tooth $RT$. The period-2 limit cycle emanating from the period doubling bifurcation $PDb$ disappears on the homoclinic curve $h3a$. In the fold point 2a two limit cycles are created, one of them undergoes homoclinic bifurcation in $h3b$ while the second annihilates in the fold point 2e with the limit cycle which appeared in the homoclinic bifurcation $h3c$.

The overall bifurcation picture can be easily understood in terms of branches of limit cycles. Using such terminology, panel (a) has four branches. The main branch links $HB$ to $LT$. The two period-2 branches connect $PDa$ to $RT$ and $PDb$ to $h3a$. Finally, the detached branch binds $h3b$ and $h3c$ together.

In each panel the rectangle indicates the region where a reconnection will take place, leading to the next panel. Parts which will form a new branch in the next panel are shown in gray. This reconnection is associated with a limit cycle cusp bifurcation point which connects two limit cycle fold bifurcation curves. Different branches appear, depending on which one of the two curves participates in the bifurcation sketch.

The transition from panel (a) to panel (b) is associated with the cusp point $A$ in Fig.~\ref{th}, and the reconnection occurs near 1b. In panel (b) the main branch therefore connects $HB$ to $h3a$ (and stays unaffected in all subsequent transitions), one of the period-2 branches connects $PDb$ to $LT$, while the two remaining branches are unchanged in this transition.

Similarly, the transition from panel (b) to panel (c) occurs near $2c$ and is associated with the cusp point $B$ in Fig.~\ref{th}. In panel (c) one of the period-2 branches connects $PDa$ to $h3b$ and the detached branch now links $RT$ to $h3c$.

The transition to the next panel (d) is associated with the cusp point $C$. The parts of the branches with $LT$ and $h3c$ are exchanged so that the detached branch directly links $LT$ to $RT$ and one of the period-2 branches connects $PDb$ to $h3c$.

The last transition to panel (e) is associated with the cusp point $D$. The link between the $LT$ and $RT$ becomes even more direct when a closed loop branch with a number of folds separates from the detached branch in the course of this transition.

The last panel (e) shows how this scenario ends. Several things happen here in no particular order: $h3b$ and $h3c$ join together; pairs of limit cycle folds (2b, 2a), (1a, 1c), (3a, 3c), (2d, 2e) and (4a, 4c) undergo limit cycle cusp or fold bifurcations and disappear; $PDa$ and $PDb$ annihilate as we move below the $PD$ curve. Also, as we reach the tip of the tooth, more closed loop branches may detach from the branch connecting $LT$ to $RT$ and disappear.

It is important to note that now both sides of the tooth, $LT$ and $RT$, are linked together and separated from any other structure. The sides are surrounded by infinitely many folds, which appear gradually on the previous panels~\footnote{Such `wiggling' is described in~\cite{Kuznetsov98} (section 6.3) and is due to the fact that a saddle-focus equilibrium, to which the limit cycle becomes homoclinic, gets more repelling than attractive. Under such circumstances, there is an infinite number of unstable limit cycles in the neighbourhood of this saddle-focus equilibrium.}.

\section{Discussion and Conclusion}

The transformation between single and double pulse excitability can be seen as the transition from the left side of the tooth $LT$ to the right side of the tooth $RT$. Panel (a) in Fig.~\ref{steps} shows that in order to follow from $LT$ to $RT$ at the base of the tooth, a period doubling bifurcation $PDa$ must be encountered. As $\D$ decreases and we move down to the tip of the tooth (shown on subsequent panels), $LT$ and $RT$ are directly linked as they are no longer connected via $PDa$. The bifurcation analysis presented here reveals the coexistence of single and double pulses on the same homoclinic bifurcation curve `crossed' by a period doubling bifurcation curve. The main mechanism which allows a continuous transition from period-1 to period-2 limit cycles (and therefore from single to double pulses) is a limit cycle cusp bifurcation.

\section{Acknowledgments}
The authors thank Dr. Dmitrii Rachinskii for stimulating discussions. This work was supported by Science Foundation Ireland under contract number SFI/01/F.1/CO13, and the Irish HEA under the PRTLI program.

\section{Appendix}
Theoretically, the dynamics of a QDL with optical injection is described by rate equations for the complex amplitude of electric field $E$, the quantum dot occupancy probability $\rho$, and the number of carriers per dot in the surrounding quantum well, $N$~\cite{Melnik06,obrien04}. The equations describing the dynamics read:
\begin{eqnarray}
\nn \dot E&=&\frac12 v_g g_0 \left( \frac{2\rho -1}{1+\eps |E|^2} - \frac{\js}{v_g  g_0}\right)(1+ \imath \al) E \\
\nn  &+& \imath E \D_m + \js \sqrt{\frac{S_m}{\hbar\w}} \nonumber, \\
\dot \rho &=&-\jd \rho + C N (1-\rho)-v_g \sigma \frac{2\rho-1}{1+\eps |E|^2} |E|^{2} \label{QDfull},\\
\nn \dot N &=&-\jn N +\frac Jq - 2 C N (1-\rho).
\label{FullModel}
\end{eqnarray}
In these equations, the parameters are (typical values given in parentheses): $\js$~$(590$~$ns^{-1})$ the photon decay rate in the cavity, $\jn$~$(1$~$ns^{-1})$ and $\jd$~$(1$~$ ns^{-1})$ the carrier decay rates in the quantum well and in the dot respectively, $C$~$(1.02$~$ ps^{-1})$ is the capture rate from the quantum well into the dot, $J$~$(6.73\times10^{-10}$~$ A)$ the current per dot, $\sigma$~$(0.6 $~$nm^{2})$ the interaction cross section of the carriers in the dot with the electric field, $\al$~$(1.2)$ the linewidth enhancement factor, $q$ the electronic charge, the detuning $\D_m$ is the angular frequency difference between the slave and master lasers, $|E|^{2}$ is the photon density, $v_g$~$(0.833\times10^{8}$~$ m/s)$ the group velocity, $\eps$~$(2\times10^{-22}$~$ m^{3})$ the gain saturation coefficient, $g_0$~$(72 $~$cm^{-1})$ the differential gain, $\w$~$(1.45$~$fs^{-1})$ the master laser frequency and $S_m$ the injected from the master laser energy density.

Let us now derive a simpler model which for the given parameter values keeps all the essential bifurcations of the full QD model, in particular a closed homoclinic tooth with a period doubling curve `crossing' it. In order to do this we first neglect the gain saturation term $\eps |E|^2$. We then note that in this regime $\rho$ can be adiabatically eliminated as
$$\rho = \frac{CN + v_g \sigma |E|^2}{CN + 2 v_g \sigma |E|^2 + \jd}. $$

The resulting reduced system is now only three-dimensional but, as required, preserves a homoclinic tooth and a period doubling `crossing' it:
\begin{eqnarray}
\nn \dot E &=& \frac12\left(\frac{v_g g_0 (C N -\jd)}{CN+2|E|^2\sigma v_g +\jd} -\js \right) (1+\imath \al )E\\
\nn &+&\js\sqrt{\frac{S_{m}}{ \hbar \w}}+\imath\D_m E,\\
\dot N &=& -\jn N +\frac{J}{q} - \frac{2CN(|E|^2\sigma v_g +\jd)}{CN + 2|E|^2\sigma v_g +\jd}. 
\label{QDreduced}
\end{eqnarray}

Further, introducing the dimensionless variables 
$$F=E\sqrt{2\sigma v_g/\jd},~M=N C/\jd,~t_1=t\js,$$ 
into Eqs.~(\ref{QDreduced}), we obtain:
\begin{eqnarray}
\nn \dot F &=& \frac12 \left(g\frac{M -1}{M + |F|^2 +1} - 1 \right) (1+\imath \al )F+\sqrt{S_{i}}+\imath\D F,\\
\dot M &=& \eta \left(- M + P - BM\frac{|F|^2 +2}{M + |F|^2 +1}\right), 
\label{SM}
\end{eqnarray}
where $g=g_0 v_g/\js$, $\eta=\jn/\js$, $B=C/\jn$, $P=J C /(q \jn \jd)$.

The Eqs.~(\ref{SM}) now have unitless time $t_1$, which is measured in units of photon cavity lifetime $\js^{-1}$, and the normalised electric field $F$, such that $|F|^2$ represents the number of photons in the cavity in the volume $2\sigma v_g/\jd$. The two new unitless control parameters are the injection strength $S_i=2S_m\sigma v_g/(\jd\hbar \w)$, and the detuning $\D = \D_m/\js$; the latter is measured in units of inverse photon lifetime $\js$. The new constants are the differential gain factor $g=1.01654$, parameter $\eta=1.695\times 10^{-3}$, parameter $B= 1.02\times 10^3$, the pump parameter $P=4.2845\times 10^3$. We now return Eqs.~(\ref{SM}) to the original notations for the phase variables, $E$ and $N$, and obtain Eqs.~(\ref{SMrev}).

\bibliographystyle{unsrt}
\bibliography{bibliography}     
\end{document}